\documentclass[twocolumn,showpacs,superscriptaddress,nofootinbib,prc,aps]{revtex4-1}

\usepackage{dcolumn}
\usepackage{amsmath}
\usepackage[normalem]{ulem}

\usepackage{xcolor}
\usepackage{multirow}

\def\nuc#1#2{\relax\ifmmode{}^{#1}{\protect\text{#2}}\else${}^{#1}$#2\fi}

\newcommand{\br}{{\vec r}}

\newcommand{\bK}{{\vec K}}

\newcommand{\pvec}[1]{\vec{#1}\mkern2mu\vphantom{#1}}

\newcommand{\be}{\begin{eqnarray}}
\newcommand{\ee}{\end{eqnarray}}

\makeatletter

\providecommand{\LyX}{L\kern-.1667em\lower.25em\hbox{Y}\kern-.125emX\@}

\usepackage{graphicx}
\usepackage{comment}
\usepackage{pifont}
\usepackage[capitalize]{cleveref}
\Crefrangeformat{Figs.}{#3#1#4-#5#2#6}

\makeatother

\begin{document}

\title{Benchmarking Faddeev and transfer-to-the-continuum calculations for $(p,pN)$ reactions}

\author{M.~G\'omez-Ramos}
\email{mgomez@ikp.tu-darmstadt.de}
\affiliation{Institut f\"ur Kernphysik, Technische Universit\"at Darmstadt, D-64289 Darmstadt, Germany}
\affiliation{Departamento de F\'{\i}sica At\'omica, Molecular y
Nuclear, Facultad de F\'{\i}sica, Universidad de Sevilla,
Apartado~1065, E-41080 Sevilla, Spain}

\author{A.~Deltuva}
\email{arnoldas.deltuva@tfai.vu.lt}
\affiliation
{Institute of Theoretical Physics and Astronomy, 
Vilnius University, Saul\.etekio al. 3, LT-10257 Vilnius, Lithuania
}

\author{A.~M.\ Moro}
\email{moro@us.es}
\affiliation{Departamento de F\'{\i}sica At\'omica, Molecular y
Nuclear, Facultad de F\'{\i}sica, Universidad de Sevilla,
Apartado~1065, E-41080 Sevilla, Spain}
\affiliation{Instituto Interuniversitario Carlos I de F\'isica Te\'orica y Computacional (iC1), Apdo.~1065, E-41080 Sevilla, Spain}

\date{\today}

\begin{abstract}
\textbf{Background}: Nucleon-knockout reactions on proton targets $(p,pN)$ have experienced a renewed interest due to the availability of inverse-kinematics experiment with exotic nuclei. Various theoretical descriptions have been used to describe these reactions, such as the Distorted-Wave Impulse Approximation (DWIA), the Faddeev-type formalism and the Transfer to the  Continuum method.

\textbf{Purpose}: Our goal is to benchmark the observables computed with the Faddeev and Transfer to the Continuum formalisms in the intermediate energy regime relevant for the experimental $(p,pn)$ and $(p,2p)$ studies.

\textbf{Method}: We analyze the $^{11}$Be$(p,pn)^{10}$Be reaction for different beam energies, binding energies and orbital quantum numbers with both formalisms to assess their agreement for different observables.

\textbf{Results}: We obtain a good agreement in all cases considered, within $\sim 10 \%$, when the input potentials are taken consistently and realistically.

\textbf{Conclusions}: The results of this work prove the consistency and accuracy of both methods, setting an indication on the degree of systematic uncertainties applicable when using them to extract spectroscopic information from $(p,pN)$ reactions.
\end{abstract}

\pacs{25.60.Gc, 24.10.Eq, 25.45.De, 25.40.Ep}

\maketitle

\vspace{2cm}
\section{Introduction}
Thanks to the  development of radioactive isotope beam facilities, experiments on unstable nuclei 
in inverse kinematics have allowed to explore nuclear structure far from the valley of stability. In particular, nucleon-knockout experiments with proton targets, $(p,pN)$ have regained popularity thanks to their simple reaction dynamics, their capacity to remove deeply-bound nucleons \citep{Ata18,Dia18,Kaw18,Hol19} and the possibility of exploring very rare isotopes using inverse kinematics. 

Their ability to explore single-particle properties for weakly- and deeply-bound nucleons render $(p,pN)$ reactions an excellent candidate to clarify the ten-year-long puzzle of the asymmetry dependence of the reduction factors ($R_s$), ratios between experimental and theoretical cross sections, for one-nucleon removal reactions. This puzzle arose from the systematic study of nucleon knockout reactions on $^9$Be and $^{12}$C targets \citep{Gad08,Tos14}, which showed a strong dependence of the $R_s$ on the difference between proton and neutron binding energies in the nucleus $\Delta S$. This dependence was not found in similar studies on transfer reactions \citep{Fla13,Kay13,Lee10}, even though very similar nuclear structure descriptions were used both for nucleon knockout and transfer, contradicting the generalized assumption that the $R_s$ factors originate from the limitations of the structure models (usually small-scale shell model), in particular, their inability to describe short-range correlations \cite{Lap93,Dic04}. This has shed doubts on the description of the reaction mechanism in these reactions, prompting careful analysis of their uncertainties \citep{Fla18,Nun11} and assumptions \citep{Baz09}.

Although very recent results of $(p,pN)$ reactions on isotopic chains for oxygen \citep{Ata18, Kaw18} and carbon \citep{Hol19} have found a small dependence of the reduction factors on $\Delta S$, in agreement with the transfer results, for these results to be reliable, the accuracy of the description of the reaction must be ascertained. At present, multiple reaction models have been used to describe the $(p,pN)$ process: the Distorted-Wave Impulse Approximation (DWIA) was extensively used in the 60s and 70s \citep{Jac66a,jac66b} and has recently been revisited  in quantum-mechanical \citep{Kaw18,Phu19} and eikonal \citep{Aum13,Ata18,Hol19} descriptions. The Faddeev-Alt-Grassberger-Sandhas (Faddeev/AGS) \citep{Cra16,Dia18,Cre19,Mec19} and  Transfer to the Continuum \citep{Mor15,Gom18,Gom17} formalisms have also been employed for the description of $(p,pN)$ reactions, using very different descriptions to the DWIA approach. Given the variety of descriptions of the $(p,pN)$ reaction, benchmarks between the different formalisms provide an useful assessment of their validity and limitations.

Following a previous benchmark between the DWIA and Transfer to the Continuum formalisms \citep{Yos18} in this work we present a systematic benchmark between Fadddeev/AGS and Transfer to the Continuum for the $^{11}$Be$(p,pn)^{10}$Be reaction, analyzing its dependence on multiple parameters, such as optical potentials, beam energy and the orbital quantum number and binding energy of the removed nucleon. The paper is structured as follows. In Sec.~\ref{sec:theory} both models are briefly introduced. The results of the benchmark are presented in Sec.~\ref{sec:results} and discussed in Sec.~\ref{sec:discuss}. Finally, Sec.~\ref{sec:summary} summarizes the main results of this work.

\section{Outline of theoretical frameworks  \label{sec:theory}}

\subsection{The TC formalism}

The transfer to the Continuum formalism \citep{Mor15} describes the process  
\begin{equation}
A + p \rightarrow C(\alpha) + p + N ,
\end{equation}
 in which an incident composite nucleus $A=C+N$  collides with a proton target, losing a nucleon (proton or neutron) and giving rise to a residual core nucleus ($C$) in some definite state $\alpha$ and  two outgoing nucleons ($p+n$ or $p+p$) by reducing it to an effective three-body problem ($p+N+C$) using the prior form of the transition amplitude:
\begin{equation}
{\cal T}^{3b}_{if}(\alpha) = \langle \Psi^{3b(-)}_{f} \phi^\alpha_C(\xi_C) |  V_{pN} + U_{pC} - U_{pA} |\phi_A(\xi_A) \chi^{(+)}_{pA}  \rangle, 
\label{T3b}
\end{equation}
where  $\Psi^{3b(-)}_{f}$ is the exact solution of the three-body problem, $\chi^{(+)}_{pA}$ is the incoming wavefunction distorted by the optical potential $U_{pA}$, $\phi_{A,C}$ are the wavefunctions of A and C respectively and $V_{pN}$ and $U_{pC}$ are the two body effective interactions between $p-N$ and $p-C$ respectively.

If the potentials are, as usual, taken to be independent of the internal  coordinates of $C$ ($\xi_C$), one can perform the integral over these internal coordinates, to give
\begin{equation}
\label{eq:overlap}
\int d\xi_C \phi^\alpha_C(\xi_C) \phi_A(\xi_A) = \sqrt{S_{\alpha,\ell,j}}\varphi^\alpha_{CA}(\br) ,
\end{equation}
where $ \sqrt{S_{\alpha,\ell,j}} \phi^\alpha_{CA}(\br)$ is an overlap wave function, with $\phi^\alpha_{CA}(\br)$ a unit normalized wave function depending on the relative coordinate of the removed particle with respect to the core and $S_{\alpha,\ell,j}$  the spectroscopic factor, resulting in a matrix element that is proportional to the spectroscopic factor:
\begin{equation}
{\cal T}^{3b}_{if}(\alpha) =  \sqrt{S_{\alpha,\ell,j}} \langle \Psi^{3b(-)}_{f} |  V_{pN} + U_{pC} - U_{pA} |\varphi^\alpha_{CA} \chi^{(+)}_{pA} \rangle .
\label{T3b_uaux}
\end{equation}
To evaluate $\Psi^{3b(-)}$, it is approximated by an expansion in terms of $p+N$ eigenstates using a discretization procedure akin to that used in the Continuum-Discretized Coupled-Channels method \cite{Aus87}, in which the final $p+N$ states are grouped ({\it binned}) in energy or momentum intervals as
\begin{align}
\label{PhiCDCC}
\Psi^{3b(-)}_{f} \approx \Psi^\mathrm{CDCC}_{f}=\sum_{n,j,\pi}  \phi^{j\pi}_{n}(k_{n},\pvec{r}') \chi_{n,j,\pi}(\bK_{n},\pvec{R}') ,
\end{align}
where $k_n$ are some  average values for the discretized $p$-$N$ energies,  $\phi^{j\pi}_{n}(k_n,\pvec{r}')$ the bin wave functions, with $j\pi$ their angular momentum and parity and  $\chi_{j,\pi}(\bK,\pvec{R}')$ are the functions describing the relative motion of the $p$+$N$ system with respect to the residual nucleus, when the former is in a given final state $\{k,j^\pi\}$. Details on the construction of these bins can be found elsewhere \cite{Aus87,TN09}. Note also that Eq.~(\ref{T3b_uaux}) resembles the transition amplitude for a transfer process, analogous to that appearing in the standard CCBA method for binary collisions \cite{Sat83}.

The angular differential cross section for a given final discretized bin can be obtained from the matrix element in a standard way, and the double differential cross section is obtained at the discretized energies $e_{pN}$ as:
\begin{align}
\label{eq:dsdw}
\left . \frac{d^2\sigma_{j,\pi}(\alpha)}{d e_{pN} d\Omega_c} \right |_{e_{pN}=
 e^n_{pN} } \simeq  \frac{1}{\Delta_n} \frac{d\sigma_{n,j,\pi}(\alpha)}{d\Omega_c} ,
\end{align}
where $\Delta_n$ is the width of the bin to which the energy $e_{pN}$ belongs. Through energy conservation, it can be readily transformed to a double differential cross section with respect to the energy of the outgoing core in the center of mass frame, and to its momentum distribution. Further details can be found in \citep{Mor15}.

\subsection{The Faddeev/AGS formalism}

Both Faddeev \cite{faddeev:60a} and 
Alt-Grassberger-Sandhas (AGS) equations \cite{AGS}
are equivalent to the Schr\"odinger equation and yield an exact description of the 
three-body system. We work with
 Faddeev equations for  transition operators in the integral AGS form 
\begin{equation}  \label{eq:Uba}
U_{\beta \alpha}  = \bar{\delta}_{\beta\alpha} \, G^{-1}_{0}  +
\sum_{\gamma=1}^3   \bar{\delta}_{\beta \gamma} \, T_{\gamma}
\, G_{0} U_{\gamma \alpha},
\end{equation}
where $ \bar{\delta}_{\beta\alpha} = 1 - \delta_{\beta\alpha}$, $E$ is the 
 energy in the three-particle
center-of-mass frame and $H_0$ is the free Hamiltonian, 
that define the free resolvent $G_0 = (E+i0-H_0)^{-1}$.
The  two-particle transition operator  is obtained from the Lippmann-Schwinger
equation
\begin{equation}
T_{\gamma} = v_{\gamma} + v_{\gamma} G_0 T_{\gamma} ,
\end{equation}
where $v_{\gamma}$ is the potential for the pair $\gamma$ in
the odd-man-out notation.
On-shell matrix elements of $U_{\beta \alpha}$, i.e.,
$\langle\psi_{\beta}|U_{\beta \alpha}|\psi_{\alpha}\rangle$,
are transition amplitudes  leading directly to the scattering observables,
with $|\psi_{\alpha}\rangle$ being the initial channel state, given by 
the product of
the bound state wave function for pair $\alpha$ and the plane wave for 
the relative motion of particle $\alpha$ and pair $\alpha$.
In the case of the breakup $\beta=0$ with the final channel function $\langle\psi_{0}|$ 
being a product of two free waves for the relative motion of three particles. The equations
are solved in the momentum-space partial-wave representation;
see Refs.~\cite{deltuva:06b,Del07} for more details.

Being a rigorous three-body formalism, the Faddeev/AGS method is more complicated and 
computationaly demanding as compared to the TC, with some additional limitations
in its applicability. First, the method of screening and renormalization for including
the Coulomb force \cite{deltuva:06b}
is restricted to two charged and one neutral particle. Furthermore,
the convergence with the screening radius and the number of partial waves
becomes slower with increasing charge;  in the present implementation 
the converged results so far are restricted to nuclei not heavier than nickel \cite{Del07}.
On the other hand,  the momentum-space integral equation framework offers more flexibility
in the employed interaction models, enabling the inclusion of 
nonlocal \cite{deltuva:09b} or energy-dependent potentials \cite{deltuva:09a}.
Thus, the Faddeev/AGS method is suitable for the inclusion of dispersive optical models \cite{dickhoff:19a},
that should be interesting to study in the future.

\section{Results \label{sec:results} }

\subsection{Application to $^{11}$Be(\protect{$p$},\protect{$pn$})}

In the following, we present the results of the benchmark calculation between Faddeev/AGS and Transfer to the Continuum (TC) applied to the reaction $^{11}$Be$(p,pn)^{10}$Be at 200 MeV/A. The masses of $^{11}$Be, $^{10}$Be, $p$ and $n$ are assumed to be respectively 11, 10, 1 and 1 average nucleon masses (1 a.n.m.~$\simeq$ 1.00797 amu). The subsystems $p$, $n$ and $^{10}$Be are taken to have spin 0 and the final breakup channels are restricted to channels where the angular momentum between $p$ and $n$ is $J=0-3$. Kinematics are computed non-relativistically to avoid ambiguities in the relativistic prescriptions used to obtain the non-relativistic Schr\"odinger equation.

The $p-^{10}$Be interaction is taken from the K\"oning-Delaroche parametrization \cite{KD03} computed at 200 MeV/A and the $p-n$ interaction binding the deuteron is the Gaussian interaction from \cite{Aus87}. The interaction between $n$ and $^{10}$Be is taken in a Woods-Saxon shape with $r_0=1.39$ fm and $a=0.52$ fm, with the depth adjusted to reproduce the binding energy of $^{11}$Be (0.5 MeV). 

In Faddeev/AGS, the $p-^{11}$Be interaction does not appear explicitly. Instead, it is  a result of the interactions between $p$, $n$ and $^{10}$Be, while in TC it is modelled through an optical potential. The choice of this optical potential adds some ambiguity to the benchmark, since no optical potential can model the $p-^{11}$Be interaction resulting from the Faddeev/AGS equations. In order to assess the effects of this ambiguity, in this work two prescriptions are presented: i) the potential is computed by folding the  $p-n$ and $p-^{10}$Be interactions over the square of the wavefunction of the bound neutron in $^{11}$Be, which in the following will be called the ``fold'' prescription, and  ii) the potential is taken from the K\"oning-Delaroche parametrization. This prescription will be referred to in the following as ``KD''.

Another difficulty in the benchmark arises from the $n-^{10}$Be interaction in the final channel. In TC, this interaction is the same for all angular momenta between $n-^{10}$Be but can be different from that binding the $^{11}$Be. However, in Faddeev/AGS calculations this interaction must be the same in the channels which have the same angular momentum and parity as the bound state, although it can be different in other channels. We explore the effect of this interaction by using two different $n-^{10}$Be interactions: i) real $n-^{10}$Be, i.e., we consider the real interaction in all $n-^{10}$Be channels, with the depth fitted to either to $2s$ or $1p$ bound state energy, depending on the studied knockout reaction. In the $2s$ case this results in also a bound $p-$wave state in Faddeev/AGS calculations, which is removed through standard methods \cite{schellingerhout:93a}, 
and a $d$-wave resonance around the relative $n-^{10}$Be energy of 1.3 MeV; ii) KD $n-^{10}$Be, i.e., we take the $n-^{10}$Be interaction from the K\"oning-Delaroche parametrization, using it for all channels in TC calculations and for all channels except the one with the bound state in Faddeev/AGS calculation. Although this results in a different treatment in TC and Faddeev/AGS, it does not generate spurious bound states and gives a more realistic absorption.

For the $p-n$ interaction, we consider two prescriptions: i) attractive $p-n$, i.e., the Gaussian potential from \cite{Aus87} reproducing the deuteron binding energy is applied for all waves; ii) parity-dependent $p-n$, i.e., the attractive $p-n$ interaction is applied for $l=0,2$ waves, while the same interaction with opposite sign is applied for $l=1,3$ waves. This prescription results in more realistic phase shifts for the $p$ wave.

In the following, calculations with real $n-^{10}$Be interaction and attractive $p-n$ will be referred to as prescription I, those with real $n-^{10}$Be interaction and parity-dependent $p-n$ as prescription II,  those with KD $n-^{10}$Be interaction and attractive $p-n$ as prescription III and those with KD $n-^{10}$Be interaction and parity-dependent $p-n$ as prescription IV. This is schematized in Table \ref{tab:prescriptions}. 

\begin{table}[]
    \centering
    \begin{tabular}{c|c|c||c|c}
    &\multicolumn{2}{c||}{$V_{n^{10}\text{Be}}$}  & \multicolumn{2}{c}{$V_{pn}$} \\
    \cline{2-5}
    \multirow{2}{*}{}     & \multirow{2}{*}{Real} &\multirow{2}{*}{KD} & $l_\text{even}$ $V_{pn}<0$ &$l_\text{even}$ $V_{pn}<0$ \\
    &&&$l_\text{odd}$ $V_{pn}<0$ &$l_\text{odd}$ $V_{pn}>0$\\
    \hline
    I& \ding{51} &&\ding{51} &\\
    \hline
    II& \ding{51} &&&\ding{51} \\
    \hline
    III&  &\ding{51}&\ding{51}& \\
    \hline
    IV&  &\ding{51}&&\ding{51} \\
    \end{tabular}
    \caption{Representation of the potentials used in the prescriptions presented in this work (See text for details).}
    \label{tab:prescriptions}
\end{table}

Results are presented for a bound neutron in $^{11}$Be in a $2s$ (1 node) and a $1p$ (no nodes) state. The observables we have chosen to explore are the transverse ($p_x$) and longitudinal ($p_z$) $^{10}$Be momentum distributions, and the $^{10}$Be energy and angular distributions in the three-body center-of-mass frame.

\subsubsection{ $2s$-state knockout}

\begin{figure}
    \includegraphics[width=\linewidth]{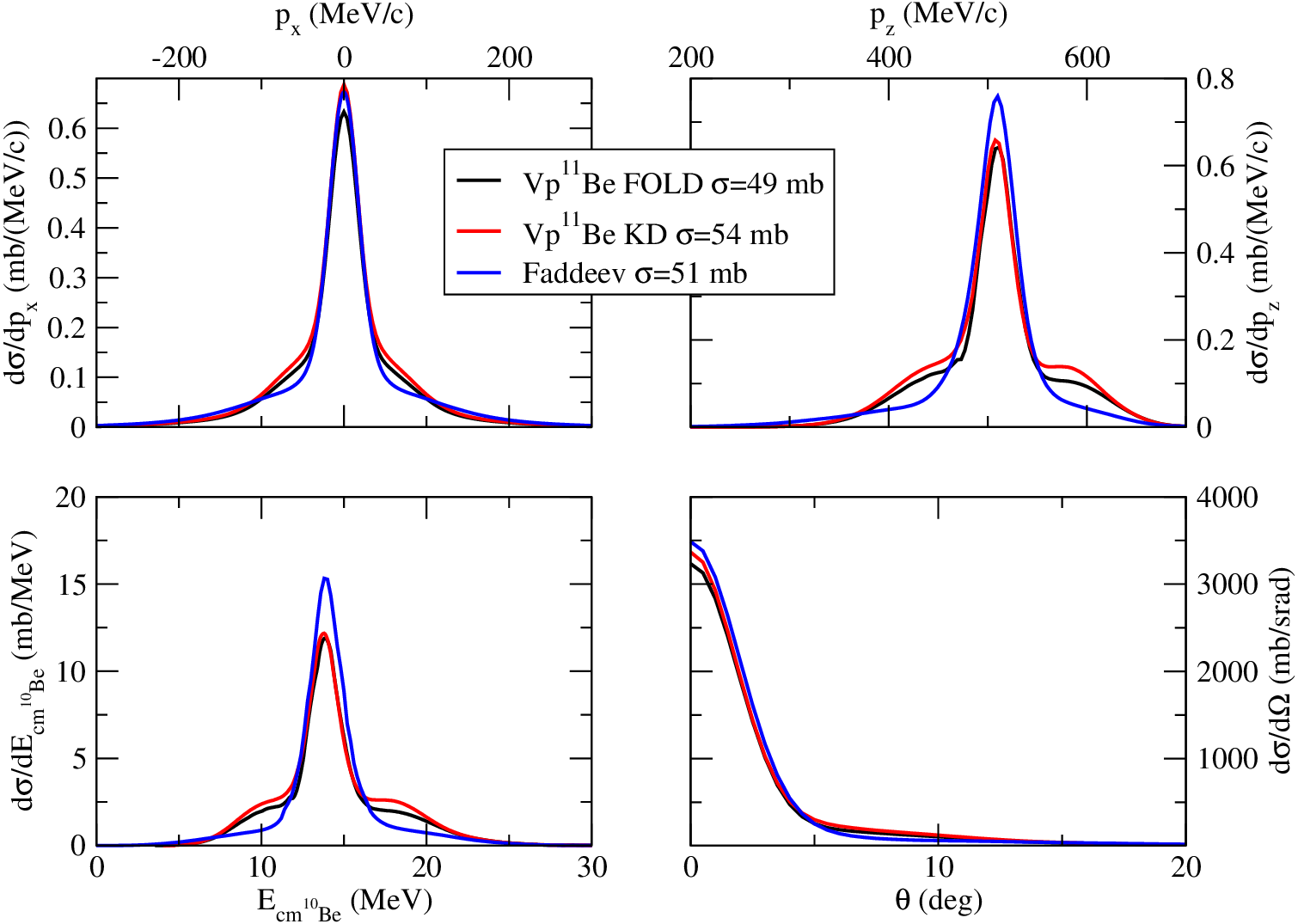}
    \caption{ Observables for the $^{11}$Be$(p,pn)^{10}$Be reaction at 200 MeV removing a $2s$ neutron with prescription I. Top left and top right correspond to the transverse and longitudinal $^{10}$Be momentum distributions, while bottom left corresponds to the $^{10}$Be center-of-mass energy distribution and bottom right to its angular distribution. 
    The black and red curves correspond to TC calculations with the incoming channel potential following the ``fold'' and ``KD'' prescriptions (See text for details).}
    \label{fig:2sbndlind}
\end{figure}

\begin{figure}
    \centering
    \includegraphics[width=\linewidth]{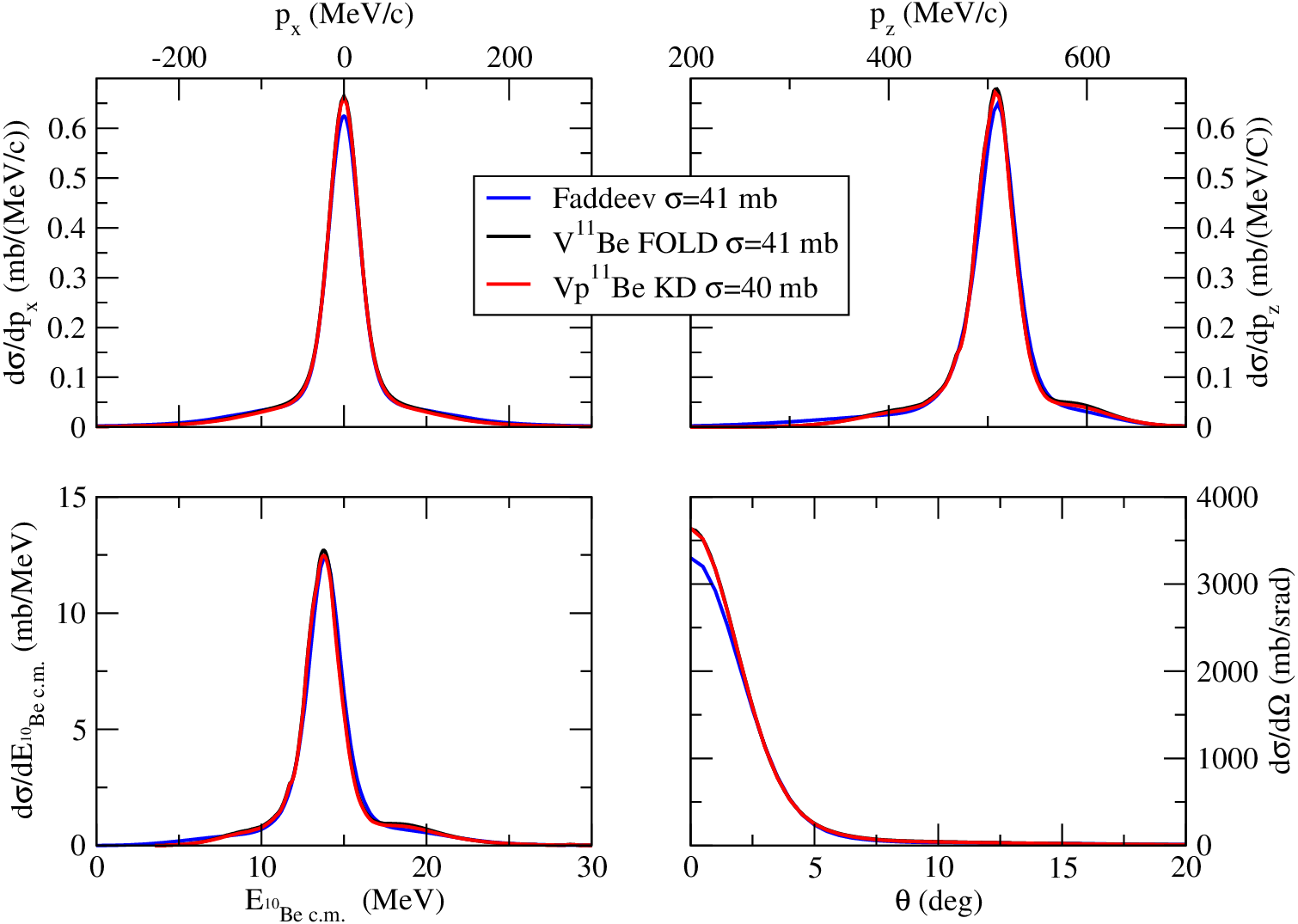}
    \caption{As Fig.~\ref{fig:2sbndlind} with prescription II (See text for details).}
    \label{fig:2sbndldep}
\end{figure}

\begin{figure}
    \centering
    \includegraphics[width=\linewidth]{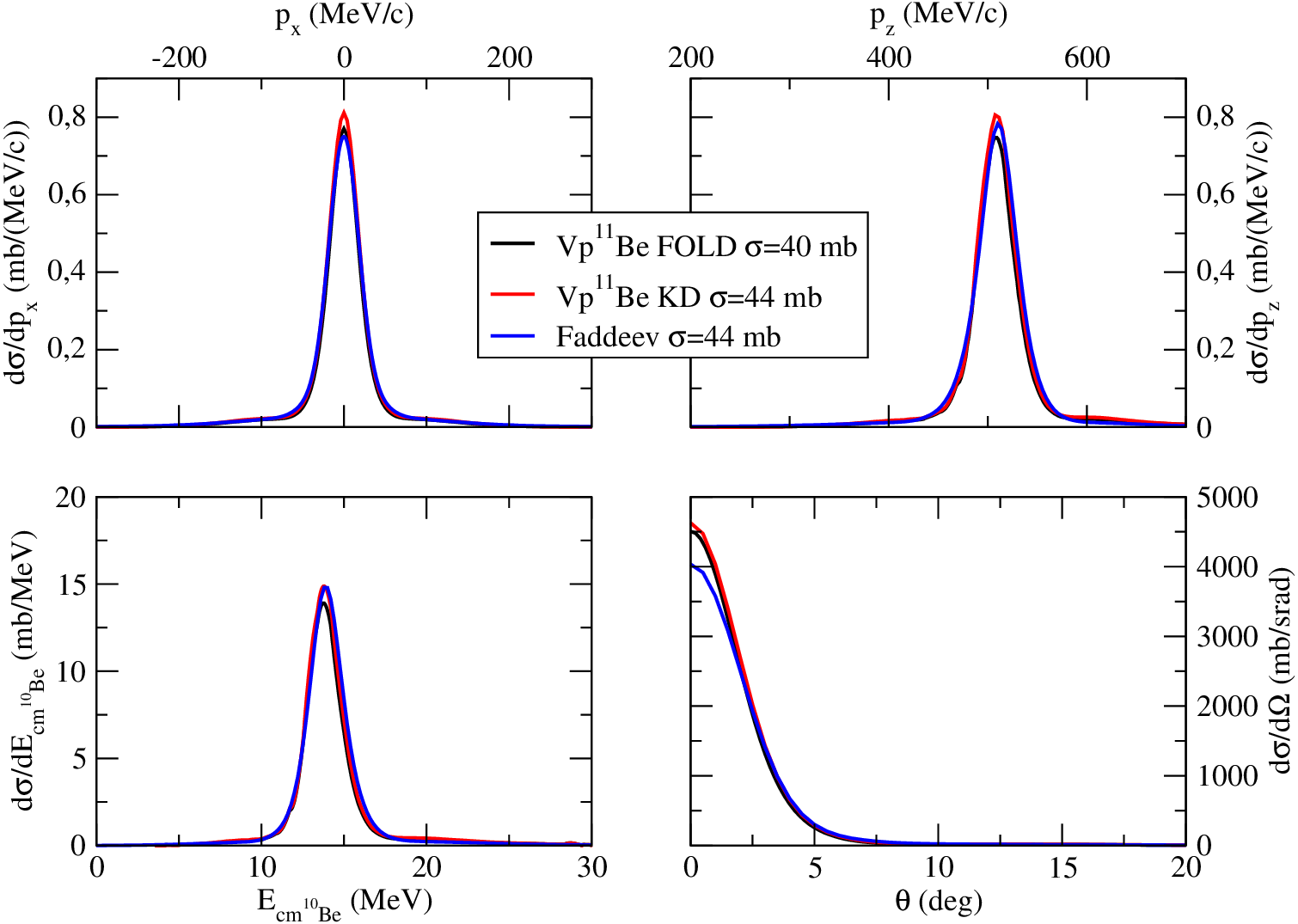}
    \caption{As Fig.~\ref{fig:2sbndlind} with prescription III (See text for details).}
    \label{fig:2skdlind}
\end{figure}

\begin{figure}
    \centering
    \includegraphics[width=\linewidth]{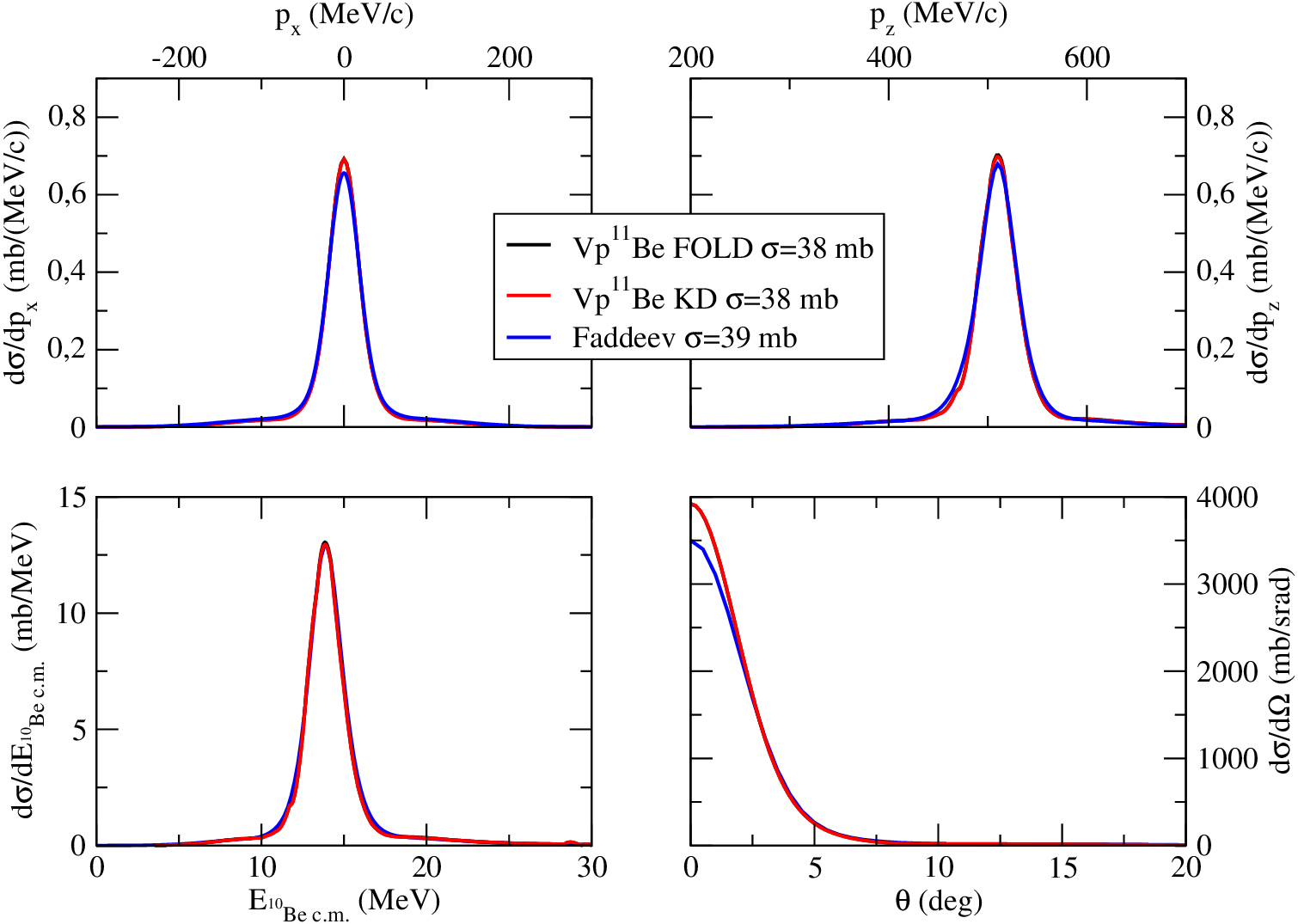}
    \caption{As Fig.~\ref{fig:2sbndlind} with prescription IV (See text for details).}
    \label{fig:2skdldep}
\end{figure}

In \crefrange{fig:2sbndlind}{fig:2skdldep}, the results of the calculations are presented for the case where the neutron is removed from a $2s$ state. 

In Fig.~\ref{fig:2sbndlind} and \ref{fig:2sbndldep} results are presented for the prescriptions I and II respectively so spurious bound states appear for the $n-^{10}$Be interaction while Figs.~\ref{fig:2skdlind} and \ref{fig:2skdldep}  correspond to the prescriptions III and IV, so they do not present spurious bound states. Among all of these, we find prescription IV to be the more realistic.

In general the agreement is good for the four prescriptions, although we find that overall, the Faddeev/AGS calculations give a consistently higher cross section, and TC overestimates the contribution at small angles.  The greatest disagreement is found for prescription I, 
where both TC calculations present shoulders in  momentum and energy distributions which are 
far less pronounced in the Faddeev/AGS calculation.

The use of the parity-dependent $p-n$ interaction leads to a better agreement between Faddeev and TC, and also between TC with the two prescriptions for the incoming optical potential. For a calculation in which the final wave function $\Psi$ is exact, the TC calculation should be independent of the choice of the incoming potential \cite{Sat83}. Therefore, we conclude that for the parity-dependent calculation, the CDCC expansion considered in TC is a better approximation of the full wave function. This could also explain the better agreement between Faddeev/AGS and TC. 

\subsubsection{ $1p$-state knockout}

\begin{figure}
    \includegraphics[width=\linewidth]{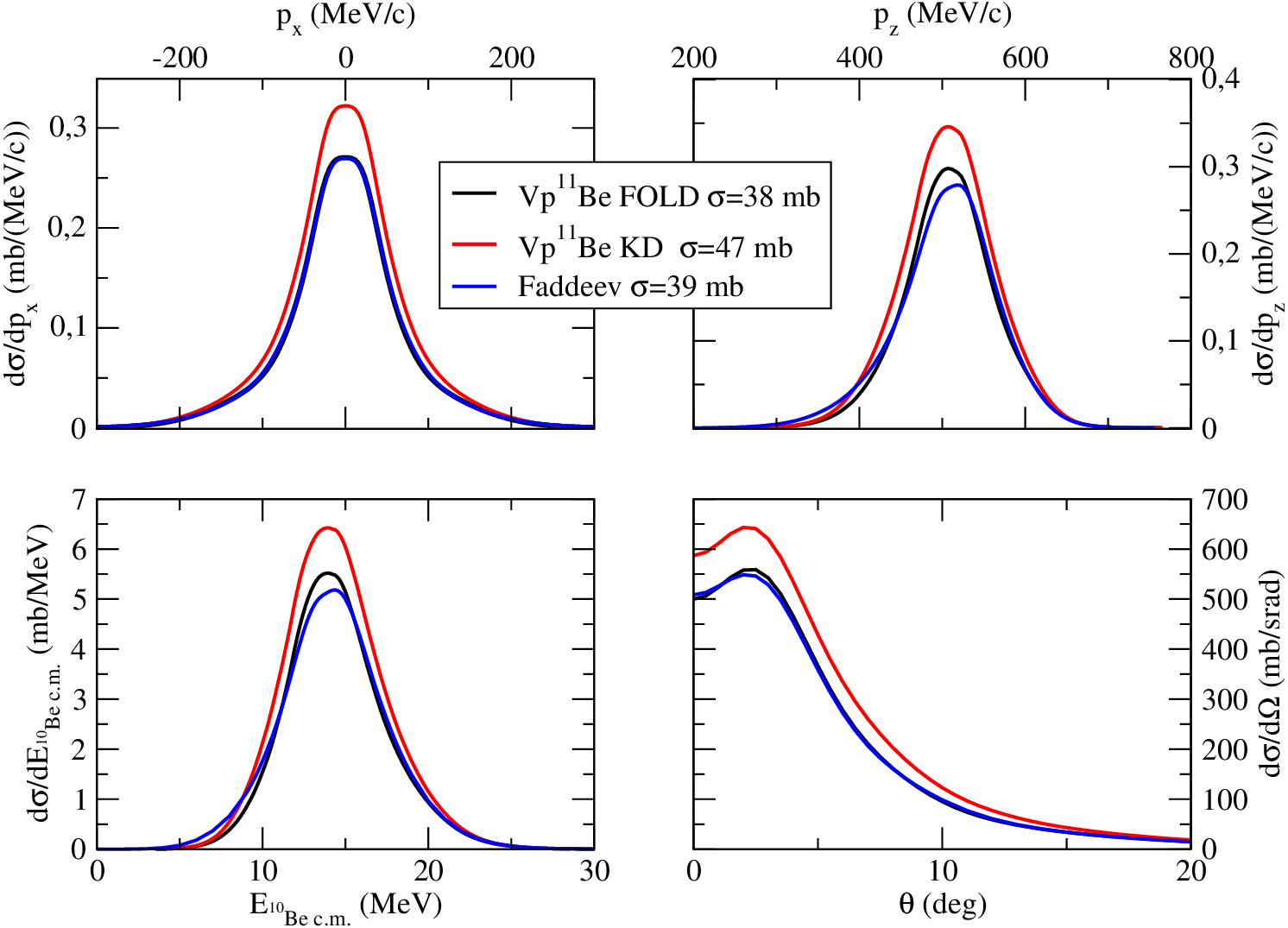}
    \caption{ As Fig.~\ref{fig:2sbndlind} for removal from a $1p$ state, with prescription I (See text for details).}
    \label{fig:1pbndlind}
\end{figure}
\begin{figure}
    \centering
    \includegraphics[width=\linewidth]{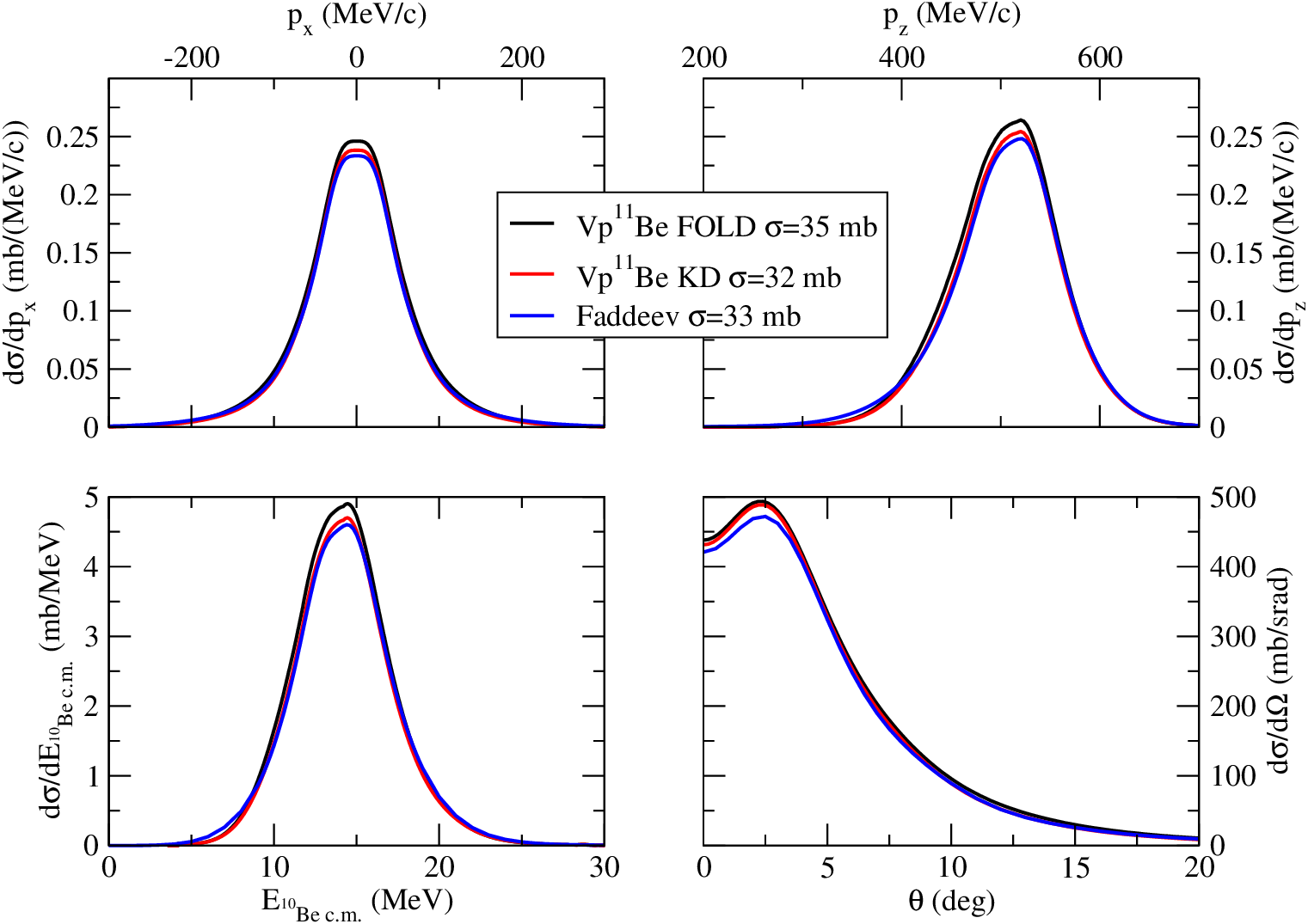}
    \caption{As Fig.~\ref{fig:2sbndlind} for removal from a $1p$ state with prescription II (See text for details).}
    \label{fig:1pbndldep}
\end{figure}
\begin{figure}
    \centering
    \includegraphics[width=\linewidth]{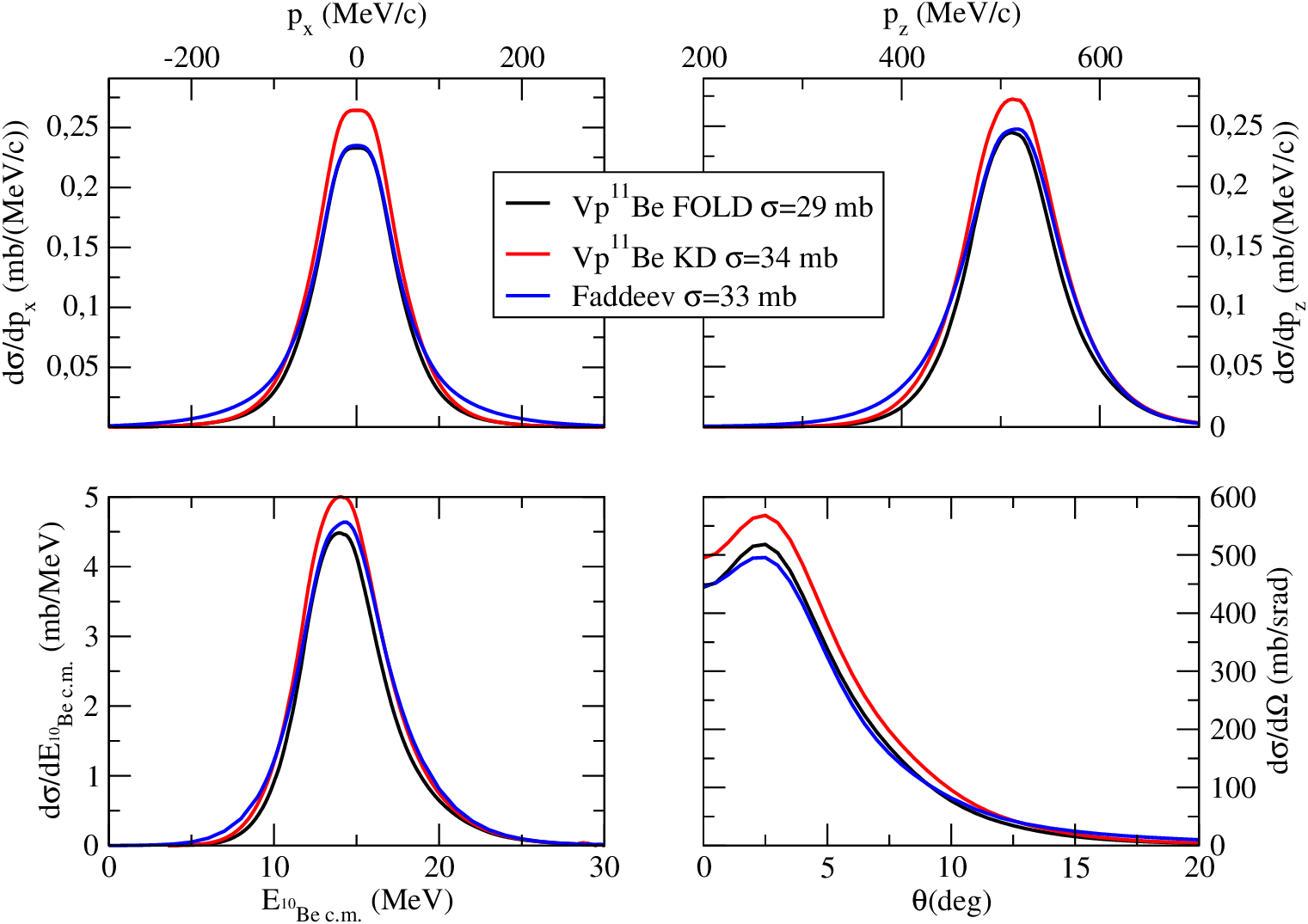}
    \caption{As Fig.~\ref{fig:2sbndlind} for removal from a $1p$ state with prescription III (See text for details).}
    \label{fig:1pkdlind}
\end{figure}
\begin{figure}
    \centering
    \includegraphics[width=\linewidth]{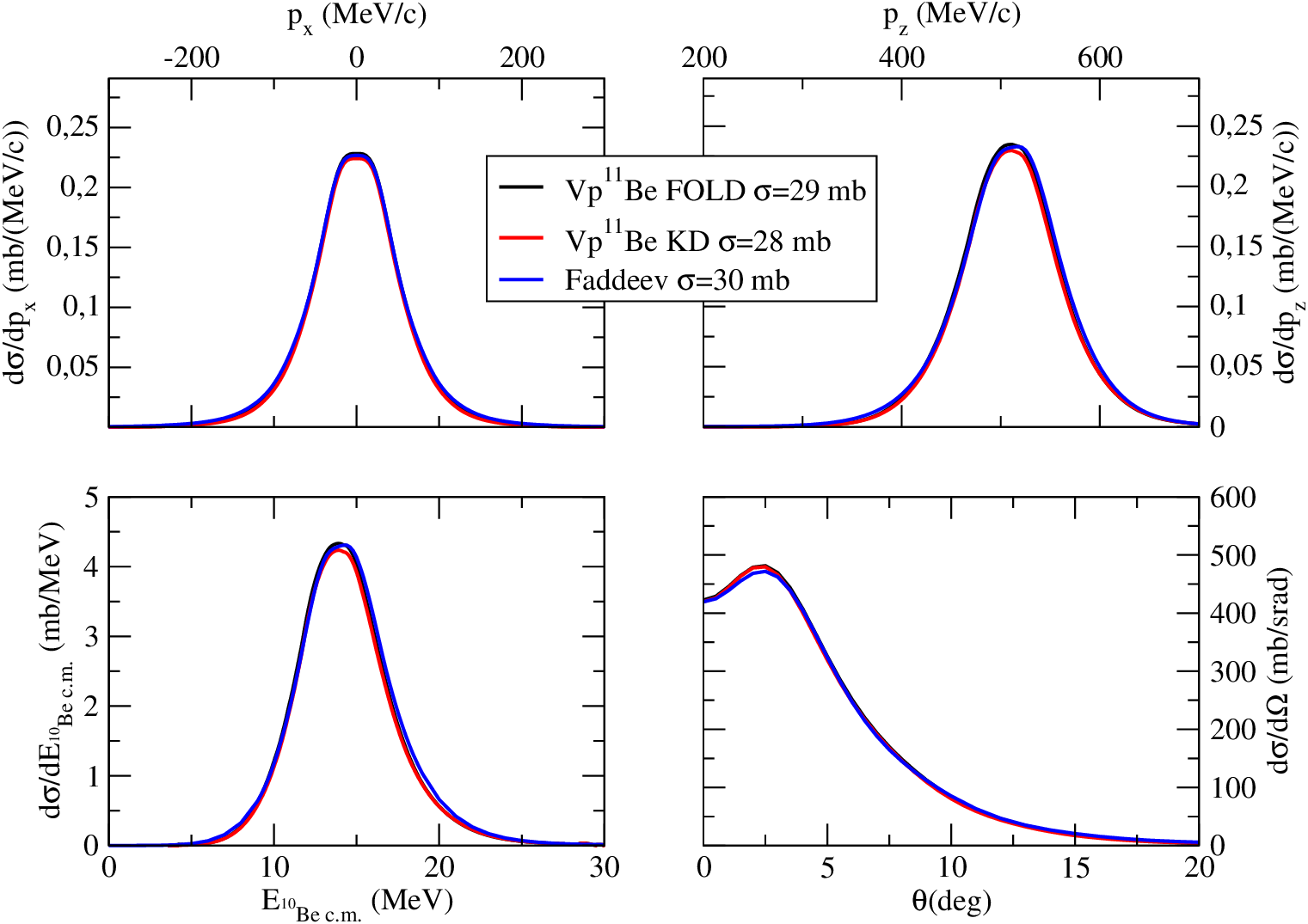}
    \caption{As Fig.~\ref{fig:2sbndlind} for removal from a $1p$ state with prescription IV (See text for details).}
    \label{fig:1pkdldep}
\end{figure}

In \crefrange{fig:1pbndlind}{fig:1pkdldep}, we present the results for the removal from a $1p$ state with prescriptions I to IV, respectively.

As in the case for 2$s$ removal, the agreement is in general good, with larger discrepancies when the fully attractive $p-n$ prescription is considered, while for the parity-dependent potential the discrepancies are lower than 10\%. For the longitudinal momentum distribution it is appreciable that the Faddeev/AGS calculations present a tail at low momenta that the TC calculation does not reproduce. This tail has been found to originate from the interaction between the neutron and $^{10}$Be, which, not being part of the prior form of the transition amplitude, is not well described in TC.

\subsubsection{ Effect of  binding energy}

As a next evaluation of our benchmark, we proceed to increase the binding energy of the removed neutron from 0.5 to 5 MeV in the calculation for $p$-wave removal, as we found it to present larger discrepancies between models. In this case, since the neutron is bound to smaller distances to the $^{10}$Be core, where absorption is more important, we find the real n+$^{10}$Be to be too unrealistic, since it does not include absorption in the outgoing $n-^{10}$Be potential. Therefore, in the following we only present results using the KD $n-^{10}$Be prescription.

\begin{figure}
    \centering
    \includegraphics[width=\linewidth]{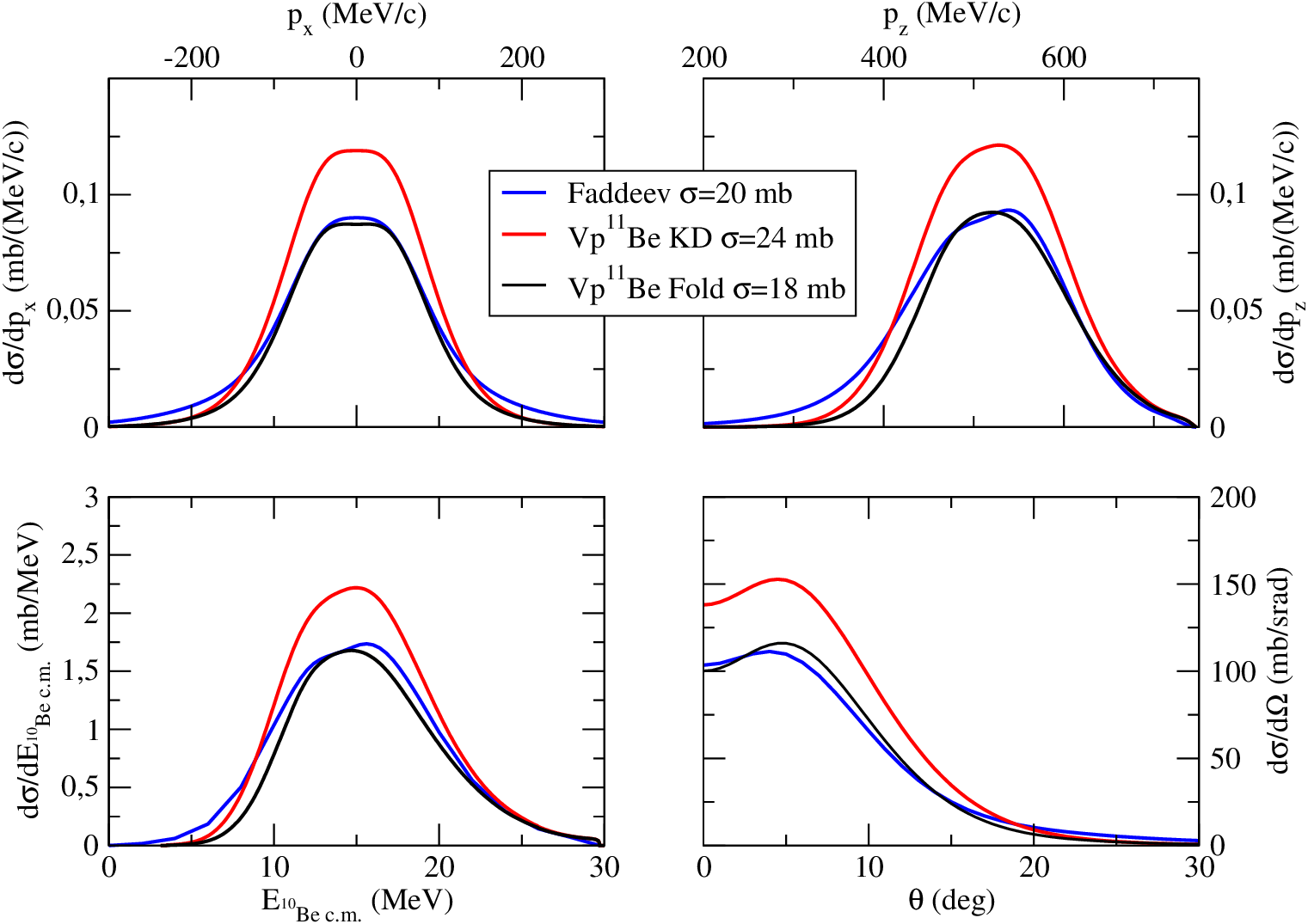}
    \caption{As Fig.~\ref{fig:2sbndlind} for removal from a $1p$ state with 5 MeV binding energy with prescription III (See text for details).}
    \label{fig:5mevlind}
\end{figure}

\begin{figure}
    \centering
    \includegraphics[width=\linewidth]{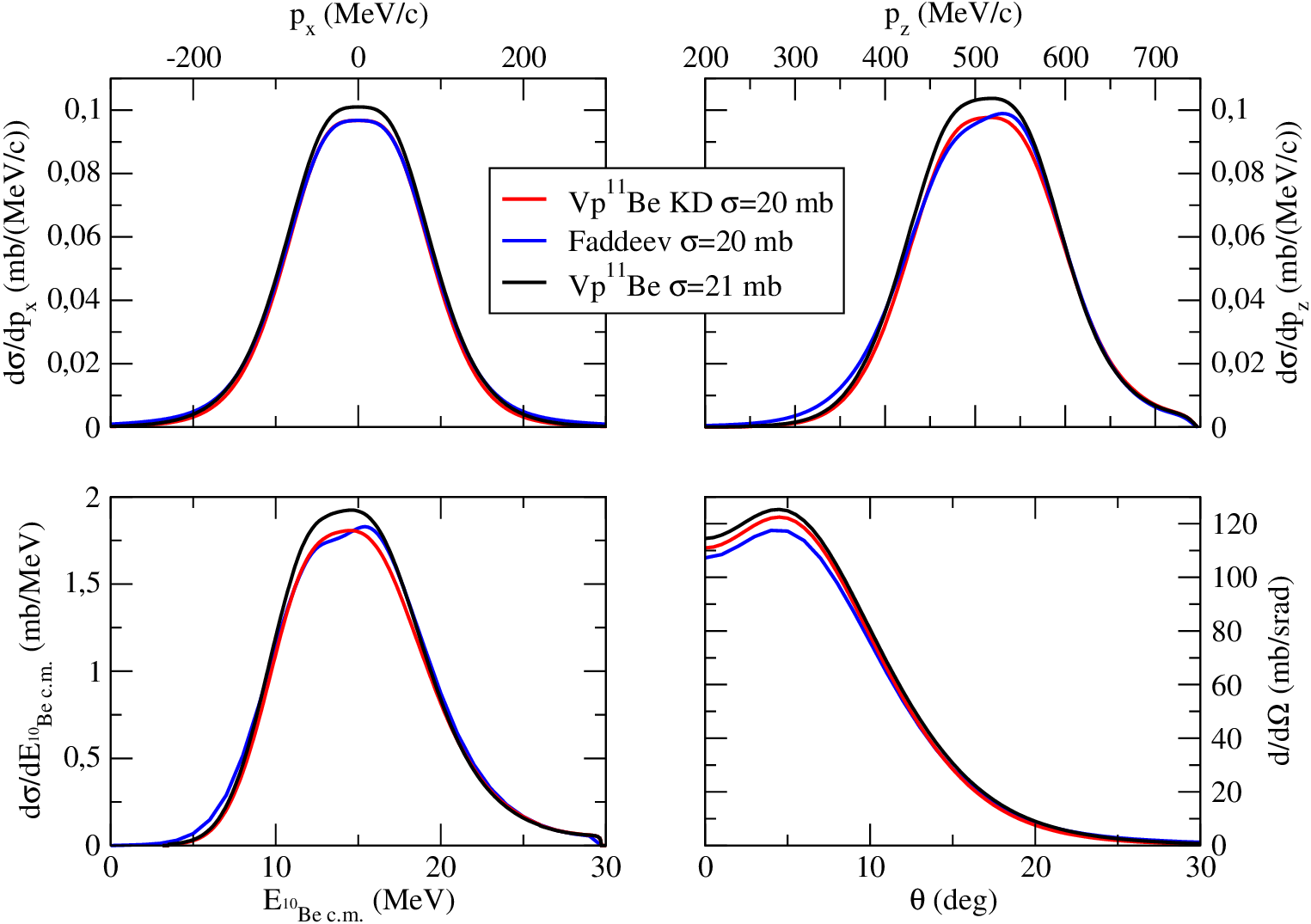}
    \caption{As Fig.~\ref{fig:2sbndlind} for removal from a $1p$ state with 5 MeV binding energy with prescription IV (See text for details).}
    \label{fig:5mevldep}
\end{figure}

Fig.~\ref{fig:5mevlind} shows the results for prescription III 
and Fig.~\ref{fig:5mevldep} shows those for prescription IV. 
It must be noted that the contribution of $(p,d)$ has been removed. In previous calculations it was negligible, but here it accounted to around 2\% of the cross section. As in the previous cases, the agreement between calculations is worse when using prescription  III, 
although it must be remarked that the differences are much more important with this larger binding energy. However, for prescription IV 
the agreement is still very good, with differences of less than 5\% between Faddeev/AGS and TC calculations, while TC calculations with different incoming potentials show larger differences than in the case with 0.5 MeV. We can interpret these results as the reaction exploring deeper parts of the wave function for a more bound nucleon, so it becomes more sensitive to the $n-^{10}$Be and $p-^{10}$Be interactions, which are described worse by the CDCC wave function, thus resulting in a larger sensitivity to the incoming optical potential. 

\subsection{Dependence on beam energy}

To test the dependence of the agreement between both formalisms on the beam energy, we present in the following two calculations for the $^{11}$Be$(p,pn)^{10}$Be reaction removing a $1p$-state neutron at 100 and 400 MeV/A. We restrict the calculations to prescription IV,
 which is the most realistic, as mentioned above. The incoming potential is taken from the KD parametrization, as that is the one giving best agreement with Faddeev/AGS using this prescription. The potentials are extracted from the K\"oning-Delaroche parametrization at the same energy as before, to isolate the effect of the beam energy from the possible change in the potentials.

\begin{figure}
    \centering
    \includegraphics[width=\linewidth]{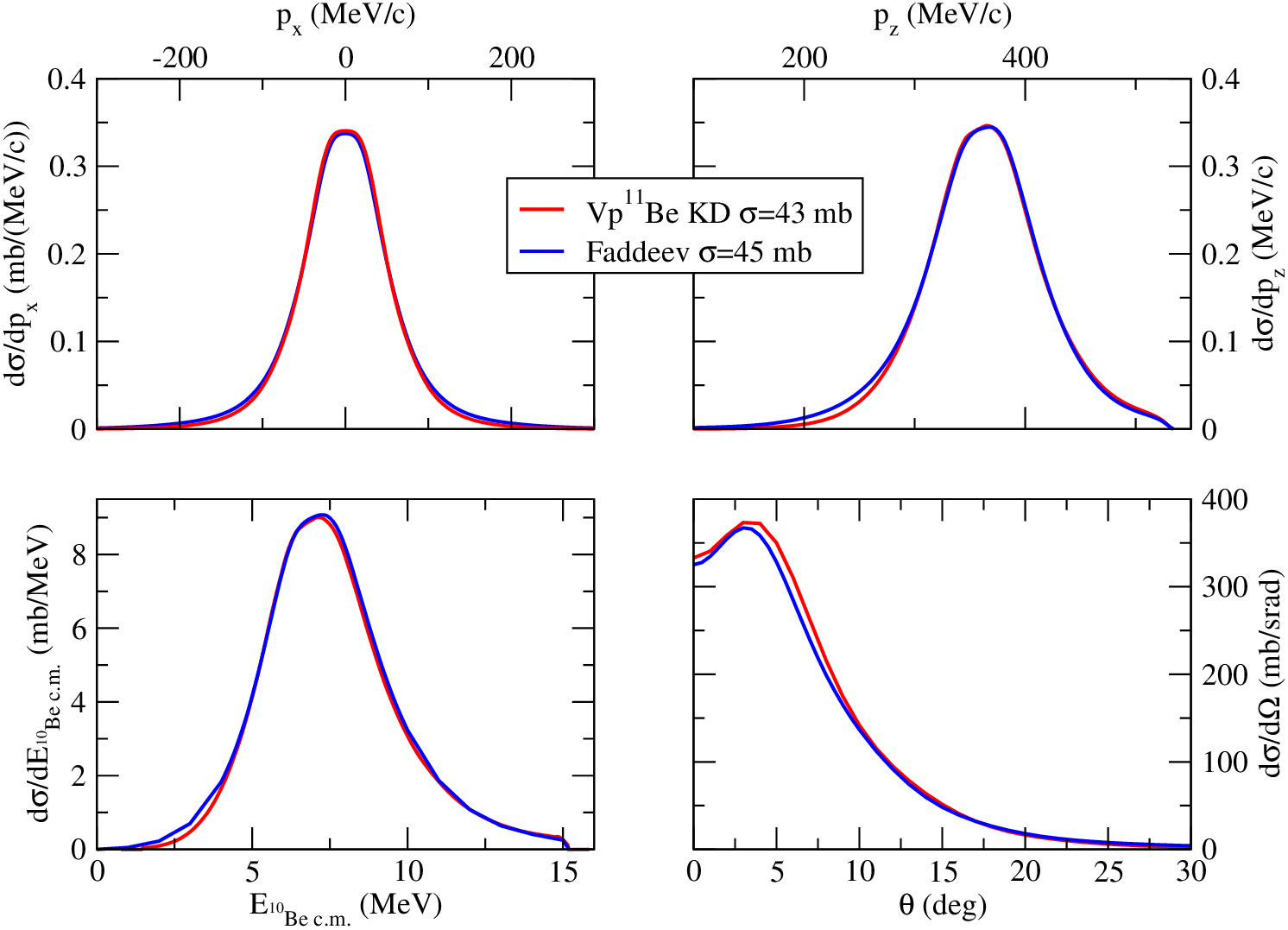}
    \caption{As Fig.~\ref{fig:2sbndlind} for removal from a $1p$ state at 100 MeV/A with prescription IV (See text for details).}
    \label{fig:100mev}
\end{figure}

\begin{figure}
    \centering
    \includegraphics[width=\linewidth]{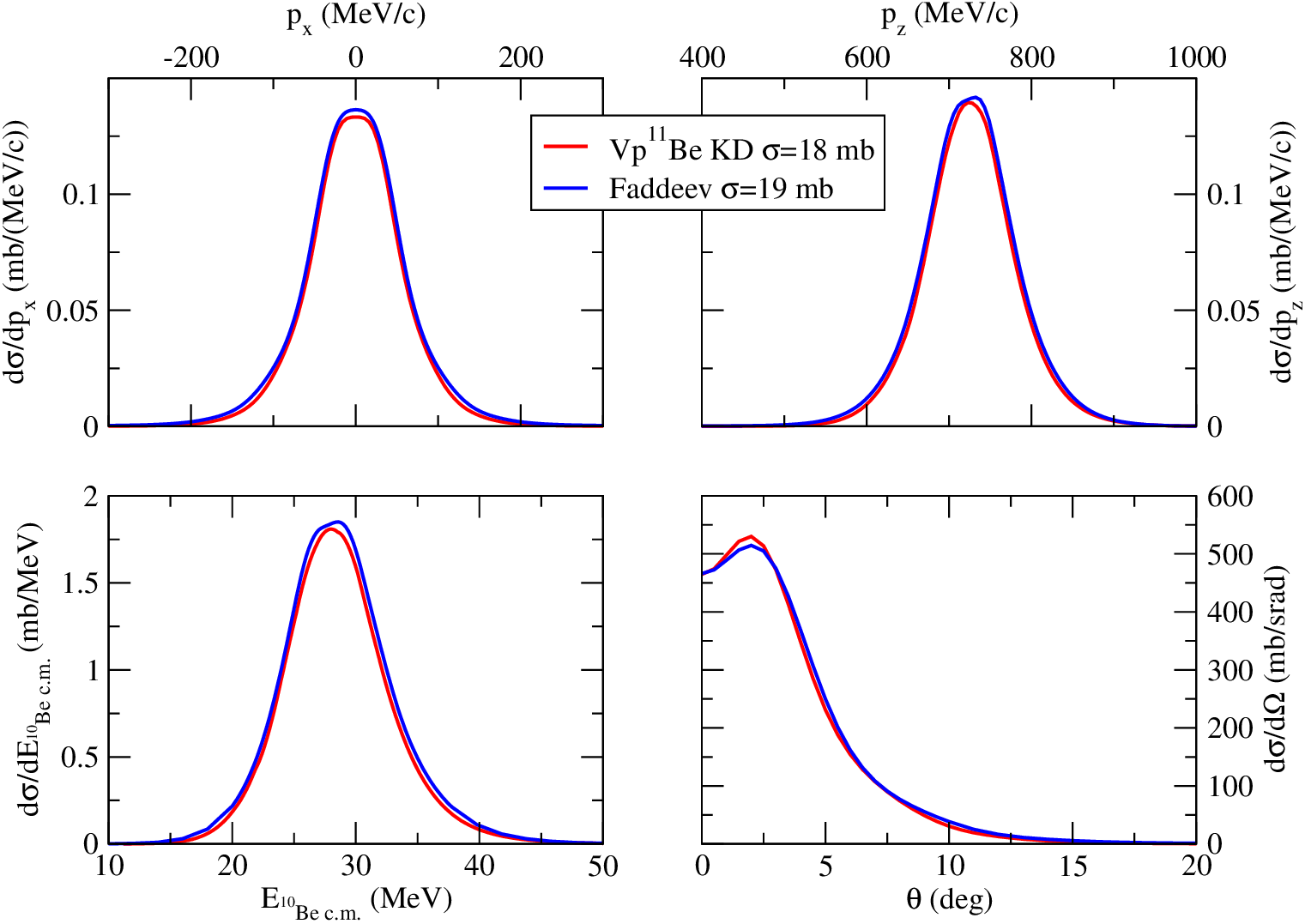}
    \caption{As Fig.~\ref{fig:2sbndlind} for removal from a $1p$ state at 400 MeV/A with prescription IV (See text for details).}
    \label{fig:400mev}
\end{figure}

Figs.~\ref{fig:100mev} and \ref{fig:400mev} correspond to the results for the reaction at 100 MeV/A and 400 MeV/A respectively. We find a very good agreement between Faddeev/AGS and TC, which seems to be rather insensitive to the beam energy. A small discrepancy can be seen in the low-energy and low-longitudinal-momentum tail at 100 MeV/A which, as mentioned before, can be related to the effect of the $n-^{10}$Be interaction.

\begin{figure}
    \centering
    \includegraphics[width=\linewidth]{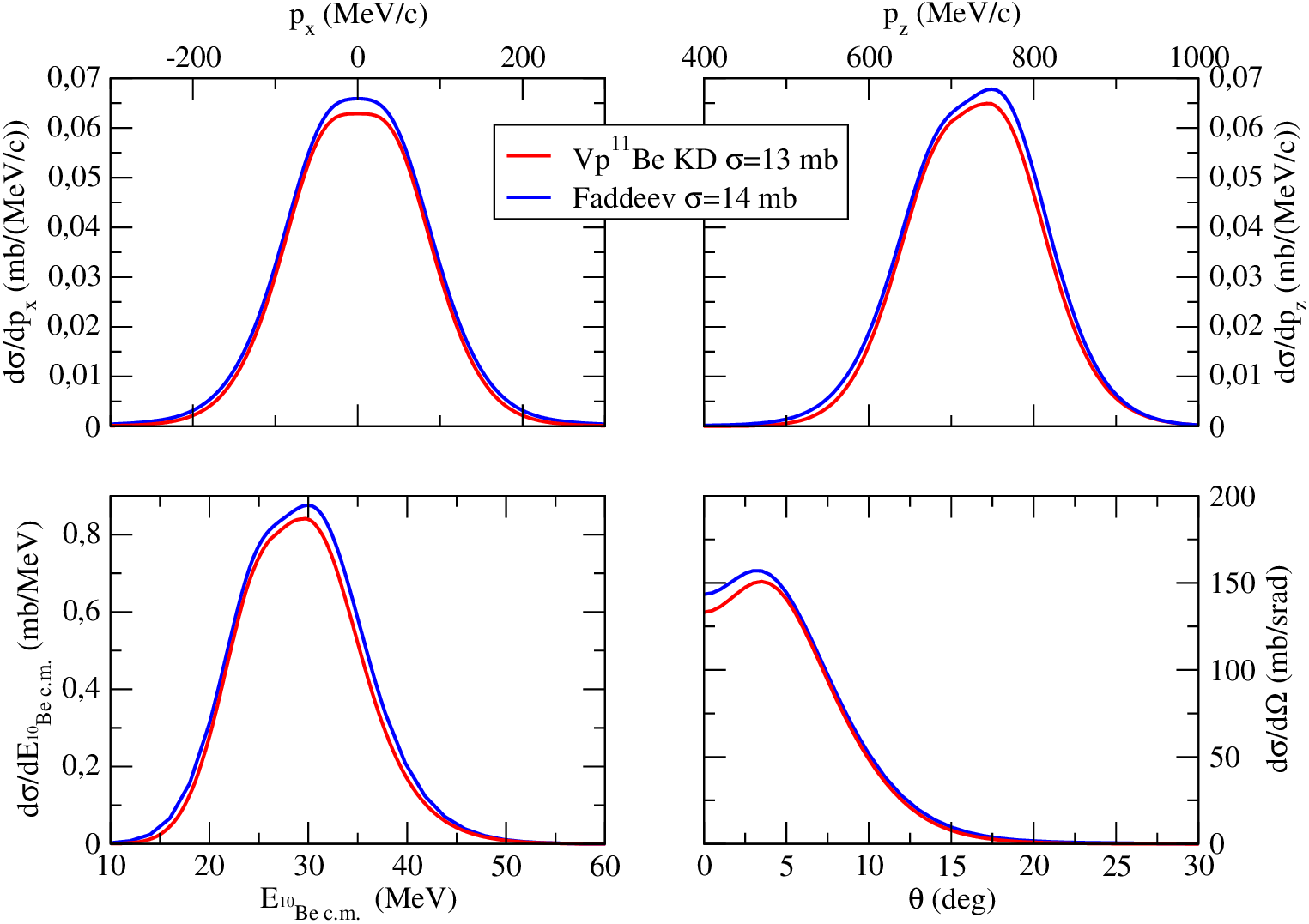}
    \caption{As Fig.~\ref{fig:2sbndlind} for removal from a $1p$ state with 5 MeV binding energy at 400 MeV/A with prescription IV (See text for details).}
    \label{fig:400mev5}
\end{figure}

In Fig.~\ref{fig:400mev5}, we present results for the removal of a $1p$ neutron with 5~MeV binding energy at 400~MeV/A. The agreement is somehow worse for the higher binding energy. We believe this worse disagreement originates from the same reasons as the previous section.

\subsection{Reid93 interaction}

\begin{figure}
    \centering
    \includegraphics[width=\linewidth]{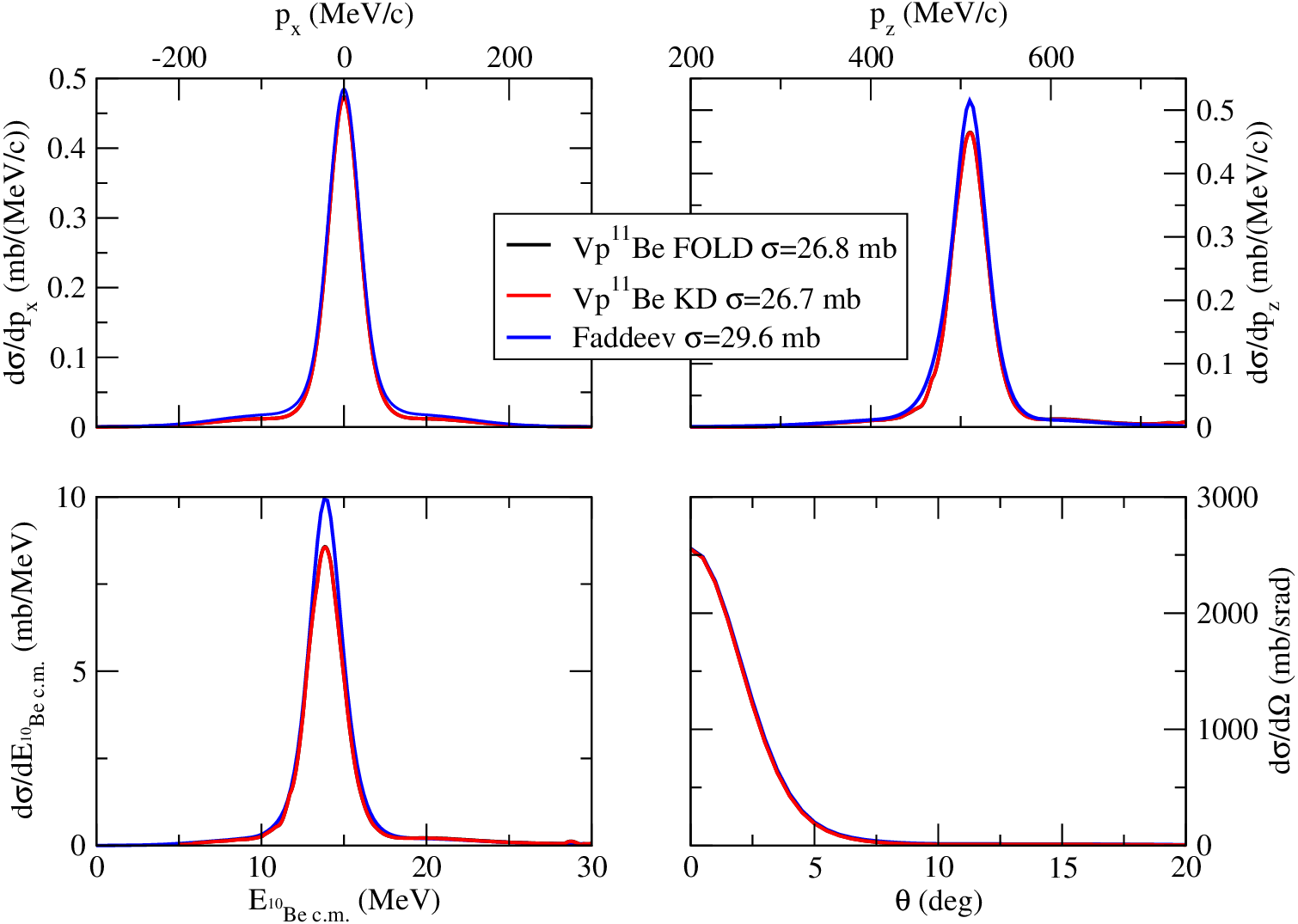}
    \caption{As Fig.~\ref{fig:2sbndlind} with the prescription KD $n-^{10}$Be and the Reid93 interaction. (See text for details).}
    \label{fig:2sRSC}
\end{figure}

\begin{figure}
    \centering
    \includegraphics[width=\linewidth]{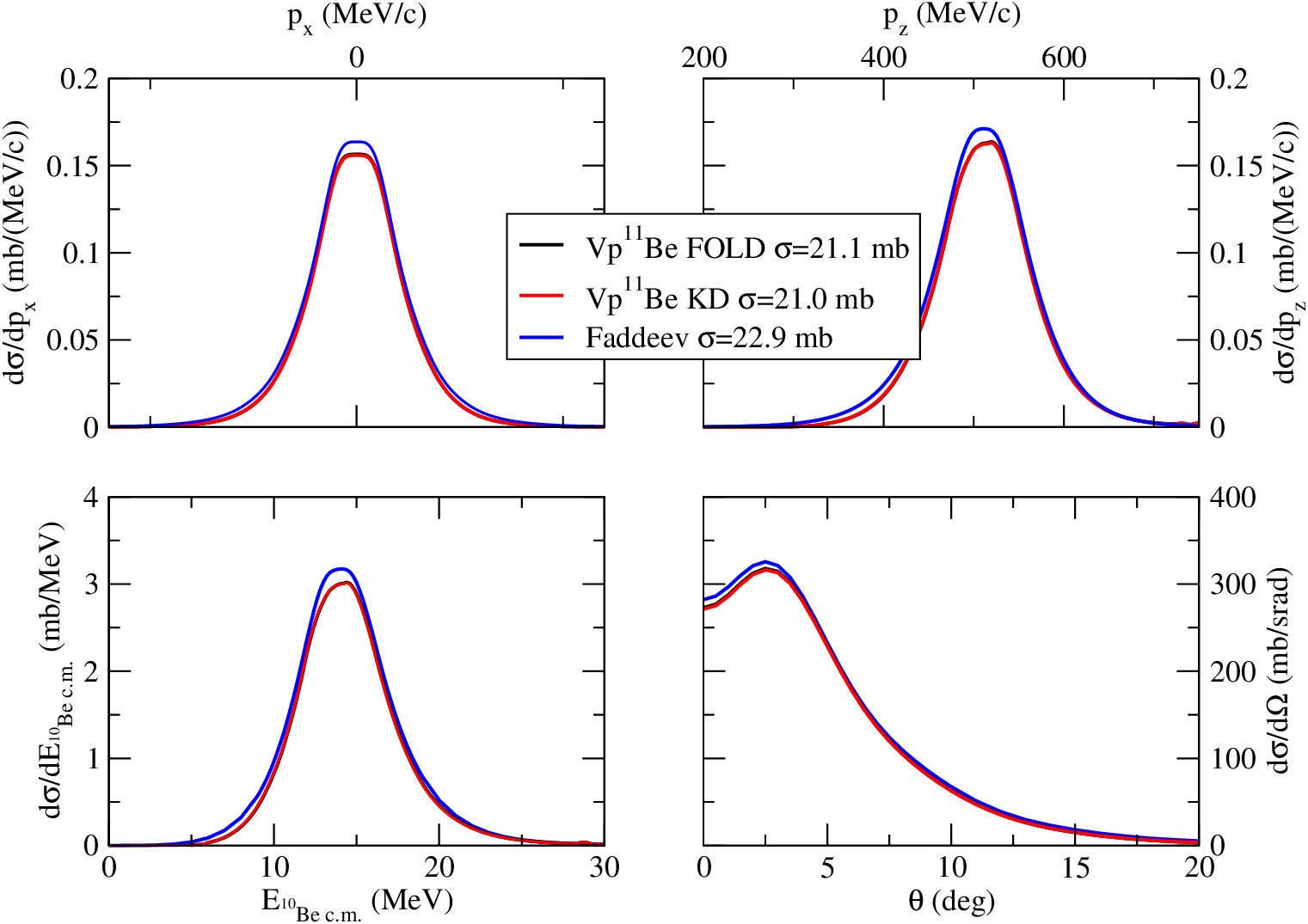}
    \caption{As Fig.~\ref{fig:2sbndlind} for the removal of a $1p$-state neutron with the prescription KD $n-^{10}$Be and the Reid93 interaction. (See text for details).}
    \label{fig:1pRSC}
\end{figure}

Given the marked differences we have found for the $p-n$ attractive and $l$-dependent prescriptions, we find it relevant to study our benchmark for a realistic nucleon-nucleon interaction. We use the Reid93 interaction \cite{Sto94}, which can reproduce the NN phase shifts with high accuracy up to energies of 350 MeV. Since this interaction includes spin-orbit and tensor terms, we consider the intrinsic spins the proton and neutron. 
Proton-neutron states are included for angular momenta $J=0-3$, including the $^3G_3$ wave, but not the $^3F_4$ one.  Due to the increase in the number of outgoing channels, in the TC calculations the outgoing channels have only been coupled to the initial channel and other outgoing channels with the same angular momentum and parity, as in previous works \cite{Mor15}. Results are presented at an energy of 200 MeV/A for the removal of a neutron with 0.5 MeV binding energy from a $2s$ state in Fig.~\ref{fig:2sRSC} and from a $1p$ state in Fig.~\ref{fig:1pRSC}. The $n-^{10}$Be interaction for the final channels is described  using the KD  potential in both cases. The incoming interaction is considered both with the ``fold'' and ``KD'' prescriptions.

We find a good agreement in this case, with differences of less than 10\% in both cases. Both incoming potential prescriptions lead to almost identical results, which points to the CDCC wave function being a good description of the three-body final state. The Faddeev/AGS results still show a larger low-energy and momentum tail, while also the peak of the distribution is larger. Test calculations using plane waves for the entrance and exit channels have been performed that show an excellent agreement, which serves as a validation of the description of the Reid93 interaction in both calculations.

\section{Discussion \label{sec:discuss}}

From the multiple cases studied in the previous section we may conclude that the TC is able to reproduce with good accuracy the more sophisticated  Faddeev/AGS results provided that the incident energy is not too low and the $p-n$ interaction is chosen realistically. For beam energies above 100~MeV/u and up to 400~MeV/u, which are typical of nucleon knockout experiments, the agreement between both methods is remarkably good, regardless of the orbital angular momentum of the removed nucleon, as $s$ and $p$-wave removal lead to similar results.
An increase in binding energy seems to lead to a moderate increase in the discrepancies between models, but, in general, the prescription for the $n-^{10}$Be interaction plays the biggest role in the agreement between formalisms.

One of the most intriguing discrepancies between TC and Faddeev/AGS corresponds to the shoulders in Fig.~\ref{fig:2sbndlind}. Tests performed by varying the $n-^{10}$Be potential in the final state show that the magnitude of the shoulders is reduced when the $n-^{10}$Be potential in the outgoing channels is set weaker. This points to their origin being related the $d$-wave resonance in the $n-^{10}$Be potential, which is not well described in TC and leads to a distortion of the final energy distribution. Fortunately, we note that realistic calculations must include imaginary components in the potentials for the outgoing particles, which severely reduce the strength of resonances, so this artifact does not appear in realistic TC calculations, such as those with prescriptions III and IV. The effect of the resonance may also explain why prescriptions I and II consistently give worse agreement than prescriptions III and IV.

The fact that the shoulders appear for prescription I but not II indicates an interplay between the $p-n$ and $n-^{10}$Be interactions in the final channels. 
This is confirmed by Faddeev/AGS calculations where the resonance peak in the
$n-^{10}$Be relative energy distribution near 1.3 MeV with the prescription I
is considerably higher than with the prescription II.
The nature of this interplay is not clear but may also be related to the better agreement for prescription IV than for prescription III. Also, it should be remarked, that for the more realistic NN interactions (prescriptions II and IV, and Reid93) the effect of this interplay does not affect TC calculations.

It is also noticeable that, in all cases, a small but visible discrepancy in the low-momentum tail is apparent between both methods. Since this tail can be attributed to the $n-^{10}$Be interaction, it cannot be well described by TC or by the commonly used Distorted Wave Impulse Approximation (DWIA) formalism. This shows the limitations of TC and DWIA, and sets a systematic uncertainty of a few percent on the observables, i.e.\ spectroscopic factors, that can be extracted using these methods.

\section{Summary and conclusions \label{sec:summary}}
In this work, the formalisms of Faddeev/AGS and Transfer to the Continuum have been benchmarked for the reaction $^{11}$Be$(p,pn)^{10}$Be at incident energies over 100 MeV/A. The benchmark shows a good agreement ($\leq$ 10\% difference), which serves as a validation of both reaction formalisms, although it must be remarked that the Transfer to the Continuum calculations lead to systematically smaller cross sections. The agreement is mostly sensitive to the nucleon-nucleon interaction between the incoming proton and the removed neutron. The use of the realistic Reid93 interaction leads to a similar agreement, of $\sim$10\%,  which is possibly smaller than the typical error in the experimental cross sections used to determine spectroscopic factors. Therefore, both formalisms can be used consistently for the extraction of spectroscopic factors, although it is essential that a realistic nucleon-nucleon interaction be used in the calculations, although the specific interaction, provided it is realistic, does not lead to significant differences \cite{Mec19}.

Given their consistency, both models can be used to extend the study of $(p,pN)$ reactions, employing each model beyond the range of applicability of the other. Namely, Transfer to the Continuum calculations show deficiencies at low energies, where the properties of the nucleon-core potentials play a significant role. There, Faddeev/AGS calculations can provide more reliable results. Conversely, for heavier nuclei and proton removal, Faddeev/AGS calculations become unfeasible, while Transfer to the Continuum can approach the computation in this regime.

It should also be mentioned that, in the presented calculation, the structure and reaction inputs (bound wavefunction and optical potentials) have been assumed to be independent. Thanks to the recent development of dispersive optical potentials \cite{dickhoff:19a} it is now possible to present a consistent analysis of the nucleon-nucleus interaction at negative and positive energies. The dispersive optical potentials present a non-local energy-dependent form. This makes them applicable in the implementation of the Faddeev/AGS equations used in this work, but not for the Transfer to the Continuum one, which would require an extension to include energy-dependent optical potentials in the coupled-channel calculations.

\begin{acknowledgments}
A.M.M. is  supported by the Spanish Ministerio de Ciencia, Innovaci\'on y Universidades and FEDER funds under project FIS2017-88410-P and RTI2018-098117-B-C21 and by the European Union's Horizon 2020 research and innovation program under Grant Agreement No.~654002. M.~G.-R. acknowledges the support of the Alexander von Humboldt Foundation.
A.D. acknowledges the support  by the Alexander von Humboldt Foundation
under Grant No. LTU-1185721-HFST-E.
\end{acknowledgments}

\bibliographystyle{apsrev4-1}
\bibliography{qfs}

\end{document}